\documentclass{an}

\usepackage{graphicx}

\usepackage{times}

\begin{document}

\headnote{}

\title{The REM Telescope: Detecting the Near Infra-Red Counterparts of
Gamma-Ray Bursts and the Prompt Behaviour of Their Optical
Continuum}

\author{F.M.Zerbi\inst{1}\fnmsep G.Chincarini\inst{1,2}\fnmsep
G.Ghisellini\inst{1} \fnmsep M.Rodon\'o\inst{3} \fnmsep G.
Tosti\inst{4} \fnmsep L.A. Antonelli\inst{5} \fnmsep P.
Conconi\inst{1} \fnmsep S. Covino\inst{1} \fnmsep G.
Cutispoto\inst{3} \fnmsep E. Molinari\inst{1} \fnmsep L.
Nicastro\inst{6} \fnmsep E. Palazzi\inst{7} C. Akerlof\inst{8}
\fnmsep Burderi\inst{5} \fnmsep S. Campana\inst{1} \fnmsep G.
Crimi\inst{1}\fnmsep J. Danzinger\inst{9} \fnmsep A. Di
Paola\inst{5} \fnmsep A. Fernandez-Soto\inst{1} \fnmsep  F.
Fiore\inst{5} \fnmsep  F. Frontera\inst{10}\fnmsep D.
Fugazza\inst{1} \fnmsep G. Gentile\inst{3} \fnmsep  P.
Goldoni\inst{11} \fnmsep G. Israel\inst{5} \fnmsep B.
Jordan\inst{12} \fnmsep D. Lorenzetti\inst{5} \fnmsep B. Mc
Breen\inst{12,13} \fnmsep E. Martinetti\inst{3} \fnmsep R.
Mazzoleni\inst{1} \fnmsep N. Masetti\inst{7} \fnmsep S.
Messina\inst{3} \fnmsep  E. Meurs\inst{13} \fnmsep A.
Monfardini\inst{14} \fnmsep G. Nucciarelli\inst{4} \fnmsep  M.
Orlandini\inst{7}\fnmsep J. Paul\inst{11} \fnmsep E. Pian\inst{9}
\fnmsep P. Saracco\inst{1} \fnmsep S. Sardone\inst{3} \fnmsep L.
Stella\inst{5} \fnmsep L.Tagliaferri\inst{1} \fnmsep
M.Tavani\inst{15} \fnmsep V. Testa\inst{5} \fnmsep F.
Vitali\inst{5}} \institute{INAF - Osservatorio Astronomico di
Brera, Via Bianchi, 46, I-23807, Merate (Lc) Italy \and
Universit\'a degli Studi Milano-Bicocca, Dipartimento di Fisica,
\and INAF - Osservatorio Astronomico di Catania, Via S.Sofia, 78 -
I-95123 Catania - Italy \and Universit\'a di Perugia, Dipartimento
di Fisica, Piazza Universit\'a 1, I- 06100 Perugia, Italy \and
INAF - Osservatorio Astronomico di Roma, Via di Frascati, 33
I-00040 Monte Porzio Catone Italy \and CNR-IFCAI, Via Ugo La Malfa
153, 90146 Palermo, Italy \and CNR-TESRE, via P. Gobetti, 101
40129 Bologna, Italy \and Randall Laboratory of Physics, 500 East
University, Ann Arbor, MI 48109-1120, USA \and INAF - Osservatorio
Astronomico di Trieste, via Tiepolo, 11 - I-34131 Trieste - Italy
\and Universit\'a di Ferrara, Dipartimento di Ingegneria, Via
Saragat 1, I-44100 Ferrara, Italy \and Service d'Astrophysique,
CEA-Saclay, 91191 Gif sur Yvette, France \and DIAS - Dunsink
Observatory, Castleknock, Dublin 15, Republic of Ireland \and
Univeristy College Dublin, Belfield, Dublin 4, Republic of Ireland
\and Univerit\'a di Trieste, Piazzale Europa, 1 I-34127 Trieste,
Italy\and CNR-IFC, via Bassini 15, I-20133 Milano}

\date{Received {\it date will be inserted by the editor};
accepted {\it date will be inserted by the editor}}

\abstract{Observations of the prompt afterglow of Gamma Ray Burst
events are unanimously considered of paramount importance for GRB
science and related cosmology. Such observations at NIR
wavelengths are even more promising allowing one to monitor high-z
Ly-$\alpha$ absorbed bursts as well as events occurring in dusty
star-forming regions. In these pages we present REM (Rapid Eye
Mount), a fully robotized fast slewing telescope equipped with a
high throughput NIR (Z', J, H, K') camera dedicated to detecting
the prompt IR afterglow. REM can discover objects at extremely
high red–shift and trigger large telescopes to observe them. The
REM telescope will simultaneously feed ROSS (REM Optical Slitless
spectrograph) via a dichroic. ROSS will intensively monitor the
prompt optical continuum of GRB afterglows. The synergy between
REM-IR cam and ROSS makes REM a powerful observing tool for any
kind of fast transient phenomena.
} \maketitle

\section{Introduction}

REM (Rapid Eye Mount) is a fully robotic fast-slewing telescope
primarily designed to follow the early phases of the afterglow of
Gamma Ray Bursts (GRB) detected by Space-borne $\gamma$-alert
systems such as HETE II, INTEGRAL AGILE, Swift. REM is currently
under construction and will be installed at la Silla Observatory
Chile, in the framework of the Fast Robotic Observatory System for
Transients (FROST) collaboration, formed by the REM/ROSS team and
the TAROT-S team. TAROT-S (see Boer et al, these proceedings) is a
fast optical imager that will provide coordinates of optical
transients with sub-arcsec precision on time-scales of a few
seconds upon trigger reception. TAROT-S is in this sense
complementary to REM/ROSS.

REM has been conceived to host a NIR (Near Infra-Red) camera
covering the 0.95-2.3 $\mu$m range with 4 filters (Z',J,H,K'). In
a second phase of the project it has been decided to host ROSS
(REM Optical Slitless Spectrograph), a slitless spectrograph
covering the range 0.45-0.95 $\mu$m with 30 sample points,
originally conceived as a stand alone telescope/instrument. As a
consequence REM will serve as a Rapid-pointing broad band
spectro-photometric facility whenever prompt multi-wavelength data
are needed.

REM is a very ambitious project since it can lead to the discovery
and study of the most distant astronomical sources ever observed
so-far. It is known that roughly a half of the GRBs observed so
far do not show any optical afterglow. At least a part of them can
be high-z bursts for which Ly-$\alpha$ absorption dumps all the
light at optical wavelengths. Ly-$\alpha$ absorption falls in the
REM wavelength range for sources with red-shift between 8 and 15,
i.e. any burst in this range can still be detected by REM and its
position determined with an accuracy of a few tenth of arcsec. The
Astrometry will be made available on a time-scale of tens of
second allowing one to observe the transient with larger area
telescopes when it is still very bright.

Via REM a 8-mt class telescope equipped with suitable IR
spectrographs, could collect a high resolution high S/N spectrum
of a source at z > 10, i.e. the most distance source within or
beyond the range of the expected red-shift of re-ionization ($8 <
z<20$). ROSS is as well an outstanding instrument since it will
intensively monitor the shape of the optical afterglow continuum
and its early temporal behaviour. ROSS will allow to constrain the
models for the prompt afterglow emission and will provide,
together with REM-IR camera, useful information about GRB
progenitors and GRB environments.

\section{Scientific Rationale}

Gamma Ray Bursts are bright, transient events in the $\gamma$-ray
sky, unpredictable in time and location, with a typical duration
of the order of seconds. The brightest bursts have $\gamma$-ray
fluences (flux integrated over the time for its duration) of $\sim
10^{-4}$ erg cm$^{-2}$. Most of the energy is released in the 0.1-
1 MeV range. Spectra generally display featureless smooth
continua.

After 30 years since their detection by the VELA satellites, we
now start to understand the physics of GRBs. This has been made
possible by the precise location of the Wide Field Camera (WFC) of
Beppo-SAX, which allowed the detection of their X-ray afterglow
emission (Costa et al. 1997) and the optical follow up
observations, leading to the discovery that they are cosmological
sources (van Paradijs et al. 1997). The huge energy and power
release required by their cosmological distances support the
fireball scenario (Cavallo \& Rees 1978; Rees \& Meszaros 1992;
Meszaros \& Rees 1993), even if we do not know yet which kind of
progenitor causes the GRB phenomenon.

The most accepted picture for the burst and afterglow emission is
the internal/external shock scenario (Rees \& Meszaros 1992; Rees
\& Meszaros 1994; Sari \& Piran 1997). According to this model,
the burst emission is due to collisions of pairs of relativistic
shells (internal shocks), while the afterglow is generated by the
shocks produced by shells interacting with the interstellar medium
(ISM, external shocks).

The emission mechanism operating during the burst proper is still
an open problem: synchrotron (Rees \& Meszaros 1994; Sari, Narayan
\& Piran 1996), quasi thermal Comptonization (Liang 1997;
Ghisellini \& Celotti 1999) and Compton drag (Lazzati et al. 2000)
have all been proposed. For the afterglow there is instead strong
evidence that the main radiation mechanism is the synchrotron
process, since this well explains the power law decay of the flux
with time, the observed power law energy spectra and the detected
linear optical polarization in GRB 990510 (Covino et al. 1999,
Wijers et al. 1999) and in GRB 990712 (Rol et al 2000). The
afterglow emission is believed to be due to the interaction of the
fireball with the circum-burst material, when enough ambient
matter is swept up by the fire-ball.

The resulting shocked material is accelerated to random
ultra-relativistic energies in an amplified magnetic field, and
radiates by the synchrotron process. Two shocks develop: a forward
shock thought to be responsible for most of the afterglow light in
all bands, and a reverse shock, thought to be responsible for the
initial emission in the IR--UV band, called the optical flash.

For about 20 GRBs we have a spectroscopic estimates of the
redshift (updated September 2001). In the case of GRB 980329 an
estimate was based on the cut-off in the spectrum interpreted as
Ly-$\alpha$ absorption by Fruchter (1999). GRB 980425 is instead
identified with the supernova SN1998bw. For those bursts of known
distance, we can calculate the energy emitted during the burst:
the energy ranges between $10^{51}$ and $10^{54}$ ergs i.e.,
assuming isotropic emission, of the order of one solar mass
entirely converted into energy.

When an afterglow is detected  the monochromatic flux decreases in
time as a power law $F_\nu (t)\sim t^{-\delta}$ with $\delta$ in
the range 0.8-2. Usually, the magnitudes of the optical afterglow
detected about one day after the $\gamma$-ray event are in the
range 19-21. It has been suggested that the steepest observed
light curves are the result of beaming. According to this idea,
the decay index $\delta \sim 1$ at early times and after a few
hours steepens to $\delta \sim 1.5 - 2.0$. Assuming m=19 after 24
hours and $\delta \sim 1.5$, the expected magnitude after 1 hour
is $\sim 13.8$. GRB 990123 has been detected by the robotic
telescope ROTSE 22 seconds after the $\gamma$-ray trigger at $m
\sim 11.7$, reaching $m\sim 8.9$ 47 seconds after the trigger
(Akerlof \& McKay 1999).

Many fascinating questions remain open about GRB origin and
structure. Some of these questions should be answered by
collecting data in the early phases of the afterglow. For these
reasons the international astronomical community developed space
borne observatories dedicated entirely (HETE II, Swift) or partly
(AGILE, INTEGRAL) to detect and send promptly a trigger with the
position of the GRB to ground-based follow-up facilities. the aim
of these facilities is to record in a timely manner relevant
information and act as an inter-mediate step to activate Target of
Opportunity procedure at telescopes with larger area. A fast
response automatic facility is then highly desirable. REM is one
of these, the only one foreseen so far to monitor the near
infrared region as well as to sample intensively the optical
continuum.

\subsection{REM and GRBs}

\begin{figure}
   \resizebox{\hsize}{!}
   {\includegraphics{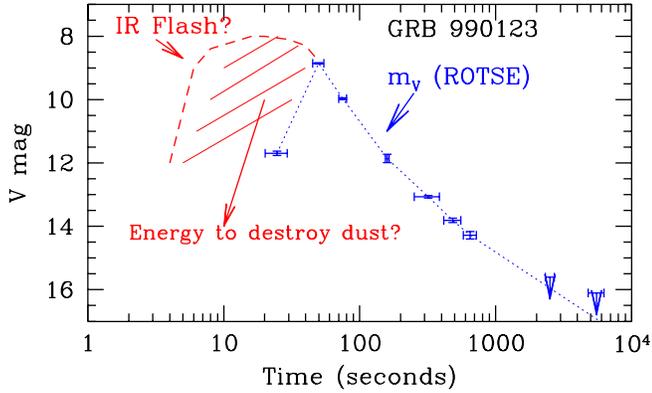}}
  \caption{The optical flash of GRB 990123 as seen by ROTSE. Part of
the optical–UV photons could have been absorbed by dust, in the
first part of the emission. After dust has evaporated, the line of
sight become extinction free. Since IR photons are much less
absorbed by dust, they could pass nearly unabsorbed, resulting in
a more prompt emission.}
  \label{fig1}
\end{figure}

\begin{figure}
   \resizebox{\hsize}{!}
   {\includegraphics{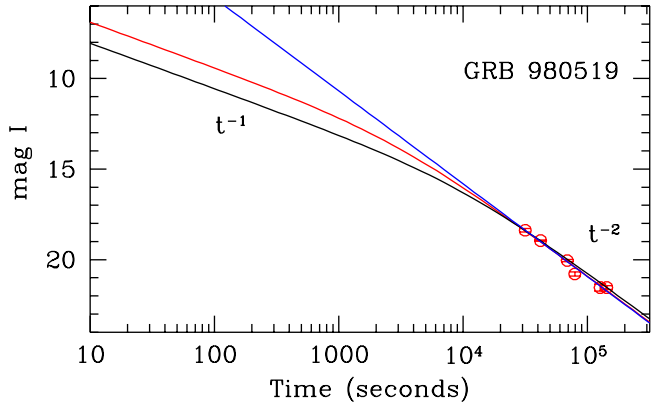}}
  \caption{The optical light curve of GRB 980519, with 3 possible extrapolation
towards earlier times. Should the early afterglow behave
"normally" (i.e. with a $t^{-1}$) we would have evidence of
beaming}
  \label{fig2}
\end{figure}

\begin{figure}
   \resizebox{\hsize}{!}
   {\includegraphics{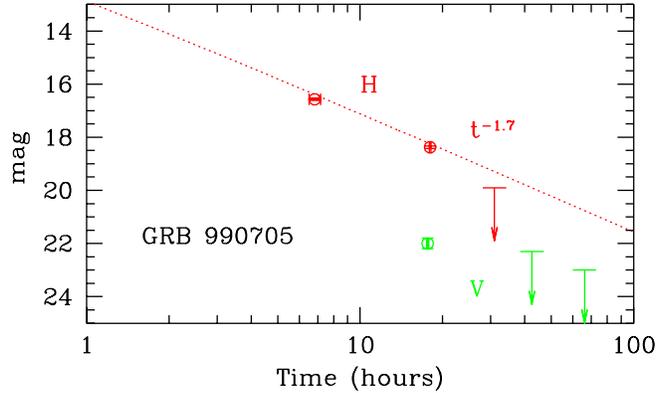}}
  \caption{The Infrared (H–band) light curve of GRB 990705, with a
possible extrapolation towards earlier times. Also shown is the
only detection point in the V–band (green point). The estimated
V-H color is 3.5-4, suggesting either a steep continuum or strong
reddening.}
  \label{fig3}
\end{figure}

Although we have only one example of prompt optical/IR emission
during the first minutes of the afterglow (GRB 990123,
figure~\ref{fig1}), we can estimate the typical early magnitudes
by extrapolating back in time the light curves of known
afterglows, as shown in figure~\ref{fig2} and figure~\ref{fig3}.
From the instruments description reported below we can see that
the REM telescope with its NIR camera is expected to reach
magnitude H=15.5, 16.04 and 17.11 with exposure times of 5, 30 and
600 seconds respectively (S/N=5). With the ROSS spectrograph (see
figures \ref{fig12} and \ref{fig13}) a V=14 point-like source is
recorded better than 10$\sigma$ in 1 sec exposures.

The above numbers suggest that we can detect the IR afterglow
during the first 2-4 hours even with an exposure time of 5
seconds. This will allow the study of the light curve and eventual
flickering in great detail, the detection of possible (even if
short) variations from the smooth power-law behavior and the
definition of any possible break. Increasing the exposure time
(after the initial phases) to 10 minutes, we can follow typical
bursts up to 12 hours, after which larger telescopes can take
over. We give below a brief description of the key points that REM
can address concerning GRBs.

\subsubsection{High-z Bursts}

If a burst is at high-z, Ly-$\alpha$ absorption dumps all the
light at optical wavelengths. Ly-$\alpha$ absorption falls in the
REM NIR camera wavelength range for sources with red-shift between
8 and 15, i.e. any burst in this range can still be detected by
REM NIR camera and its position determined with an accuracy of a
few tenths of an arcsec. The simultaneous detection in the IR and
a non-detection in the optical can directly flag the presence of a
high-z object.

A highly absorbed burst might as well dump the optical photons out
but the color-color techniques can discriminate between the two
possibilities and select good candidates high-z objects
automatically and in real time. Large telescopes (such as VLT) can
then point at the target while it is still bright enough for
high-dispersion spectroscopic observations. This appears to be the
only way to obtain high quality spectra for large redshift (even
$z>10$) objects, to study the distribution of Ly-$\alpha$ clouds
with redshift, their metallicity etc. Moreover, in the majority of
cases, GRBs are associated with galaxies. The detection of a very
high redshift GRB would hence mark the likely location of a very
high redshift galaxy.

When REM will be operated with Swift the information of the non
detection of an optical transient will be promptly made available
by the UVOT monitor on-board the satellite. When operated with
other trigger systems, however, the collaboration with TAROT-S in
the framework of the FROST Collaboration  will be of paramount
importance.

\subsubsection{Reddened Bursts}

If the burst is not at high redshift the absence of an optical
afterglow can be due to absorption by intervening matter (dust),
either in the close vicinity of the burst (if exploded in a dense
star forming region whose dust has not been completely destroyed
by the burst emission itself), or by dust distributed along the
line of sight, even at large distances from the burst site. In
this case the infrared light is much less absorbed, and therefore
an IR transient can be detected even if the optical is not. From
this point of view, the REM telescope, combining the two (IR and
optical) datasets, will makes possible to estimate the amount of
absorption.

The above is important since a lack of dust absorption,  thought
to be associated with star-forming regions, has been reported in
the spectra of bursts.  A possible explanation for such a lack is
that dust grains are sublimated by the prompt optical/UV emission
(see above and Fig. 1). In this case, an {\em IR flash} should be
observed to start before the optical one, as the IR radiation can
penetrate unabsorbed in the cloud while the higher energy photons
progressively clean out the dust. The observed IR fluence before
the detection of the optical flash would greatly constrain the
amount of dust in the cloud.

\subsubsection{Broad Band Spectrum}

Even if the theory of afterglow emission  (due to synchrotron) is
largely accepted, the best constraints we have come from
observations performed one or several days after the GRB. In fact,
the polarization was observed $\sim$ 1 day after the explosion,
and, for example, the broad band spectrum of GRB 970508 has been
measured 12 days after the $\gamma$-ray trigger. It is however
possible (if not even likely) that at the beginning of the
afterglow the physical conditions (density and temperature) of the
shocked material are more extreme and that different emission
processes (e.g. Comptonization) are dominant. A non-uniform ISM
distribution (e.g. a wind r$^{-2}$ profile) would strengthen these
differences even more. Increasing the "frequency leverage"  is
therefore of great importance to better determine the spectrum of
the early afterglow. The available information from a good quality
spectrum is a test of the afterglow emission process, a constraint
of the ISM distribution (and hence gives insight for the problem
of the progenitor) and is a measure of the optical extinction.

\subsubsection{Same Co-moving light curve}

If bursts with different redshifts are observed at the same
wavelength, a comparative analysis is hardly possible since the
rest frame wavelength of the observations is different. If,
however, multifilter observations are systematically performed for
all bursts, it is possible to compare between them magnitudes (or
fluxes, spectra) at the same comoving frequency. The larger the
spectral baseline of the observations, the larger set of redshifts
can be consistently compared. The wide spectral range covered by
REM-IR camera and ROSS spectrograph allows the comparison in an
unprecedented wide range of redshifts.

\subsection{Additional Science Program}

The REM telescope, as all other robotic facilities dedicated to
GRB science, reacts to a trigger from a space-borne satellite.
This means that for a considerable amount of time REM will be idle
in the sense that it will not be pointing at any GRB transient.
Such a time depends on the number of public triggers eventually
provided by missions scheduled to fly during REM operation is
estimated to be around 40\% of total REM observing time. During
the idle phase REM will serve the community as a fast pointing NIR
imager and optical spectrograph particularly suitable for
multi-frequency monitoring of highly variable and transient
sources.

Among the obvious applications of REM idle time there are AGN and
variable star multi-frequency monitoring. Some Key-programs of
interest have been already identified by the REM science team and
the related preparatory work has been initiated. REM will be used
in association with INTEGRAL to monitor Galactic Black Hole
candidates flare stars and with AGILE for Blazar monitoring.

\subsubsection{Galactic Black Holes Candidates}

To obtain a complete picture of accreting binary systems, near
infrared and optical observations can be fundamental as they allow
one to estimate the compact object mass through mass function
measurements. Indeed this is the best method available to
establish the real nature of Black Hole Candidates (BHCs)
identified at other wavelengths (usually X-rays). Even if this
measure is not always possible, other measurements of e.g. the
spectral type of the companion star are of great help in
understanding the accretion phenomenon and the evolutionary
history of the system. Near Infrared (NIR) observations are better
suited to this research than optical ones as they suffer less from
contamination from the accretion disk and galactic absorption.
Optical data can instead help in the determination of the spectral
type of the companion.

Of particular importance in this field are the observations of
X-ray Novae (Tanaka \& Shibazaki 1996), this class of transient in
fact contains the majority (11 of 14) of BHCs with dynamically
measured mass functions (Tanaka \& Shibazaki 1996, Filippenko et
al, 1999, Mc Clintock et al. 2001, Orosz et al. 2001). These
sources are discovered during an outburst during which they first
become very bright, then progressively less luminous and in the
end almost undetectable. Their X-ray emission evolves into
different spectral states which probably correspond to different
mass accretion rates. Most of all, at the end of the outburst
X-ray emission becomes so faint that it is possible to measure the
optical mass function of the system and hence the BH Mass. X-ray
Novae offer therefore an ideal test case for the study of
accretion processes in Black Holes. Several phenomena at different
wavelengths (radio jet emission, optical/NIR brightening...) are
often visible during the X-ray Nova outburst.

To properly characterize the outburst mechanism and the succession
of events, a fast identification of the optical counterpart is
essential in order to allow a continuous monitoring parallel to
X-ray results. Given the relatively bright magnitudes of these
objects ($V \sim 13$ at maximum for Nova Muscae 1991 (Della Valle
et al. 1991)) and the error box that can be provided by
present-day X$\gamma$-ray instruments (less than 1 arcminute), REM
can easily perform the identification and follow the light curve
with multicolor photometry (and slitless spectroscopy) while high
signal to noise spectra could be obtained with larger telescopes.

\subsubsection{Blazars}

Blazars are the most violent subclass of active galactic nuclei,
emitting across the entire electromagnetic spectrum, and being
characterized by large amplitude variability in short time-scales
(factor 2 in hours, factor 100 in weeks–months). Blazars are
important because their emission is produced by plasma flowing at
relativistic speeds through jets, from the very inner core of the
nucleus (i.e. a few Schwarzchild radii) to distances as great as
hundreds of Kpc.

The spectral energy distribution (SED) of Blazars is characterized
by two broad peaks: the first is in the IR–UV band, the second the
MeV–GeV band. There is unanimous consensus that the first broad
peak is due to synchrotron emission, while the nature of the
second peak is still under discussion. Most likely, it is due to
the inverse Compton process by the same electrons producing the
synchrotron photons.

Recent multiwavelength observing campaigns seem to find that a
coordinated variability at different frequency bands is present in
Blazars. This would mean that the two peaks of emission are caused
by the same population of electrons at the same location of the
jet and that there is a localized region, in the jet, where most
of the radiation is produced. The same campaigns point out that at
frequencies lower than the peak the variability is less violent.

This could be due to a difference in cooling times between the
electrons producing the higher frequency and the lower frequency
flux. It could be also due to a different dimension of the portion
of the jet corresponding with the flux formation in the sense that
the flux at lower frequencies is produced in a more extended part
of the jet. The way to discriminate between these two options is
dense monitoring at frequencies below and above the synchrotron
peak, to search for any possible time lag (which would favor the
first hypothesis) or simultaneous variations of decreasing
amplitude for decreasing frequencies (which would favor the second
hypothesis).

Blazars are classified according to the presence or absence of
strong emission lines and according to the location of the
synchrotron peak. Fossati et al. (1998) and Ghisellini et al.
(1998) showed that the different behaviour depends mainly on the
bolometric luminosity of the source. For instance Blazars peaking
in the near infrared are the subclass of the classical BL Lacs and
FSRQs discovered through radio means. We then expect that the IR
spectrum for these sources is characterized by a power law of
index $\sim 1$: $F(\nu) \propto \nu^{-1}$. Steeper spectra are
expected for more powerful sources (having their synchrotron peak
at even lower frequencies), and flatter spectra for lower
luminosity BL Lacs.

The limiting sensitivity of REM-IR camera and ROSS spectrograph
allows one to observe about 2/3 of all known blazars (i.e around
300 objects). This means that there will be plenty of possible
candidates for observations with REM. In addition, since REM will
be operational at the same time as AGILE (sensitive between 0.05
and 30 GeV), simultaneous observing campaigns can be organized, or
prompt ToO REM observations performed in response to particularly
in-tense flaring $\gamma$-ray states of some Blazars.

\subsubsection{Flare Stars}

Flare stars may produce $\gamma$-ray fluxes of the same order as
GRBs. Flare events occur in the atmospheres of several types of
stars, from pre-MS to post-MS cool stars and involve the entire
stellar atmosphere, from the photosphere up to coronal layers.
Owing to the rather different physical characteristics of these
different layers, with temperatures and densities spanning several
decades, the flare occurs in a wide range of wavelengths, from
microwaves to X-rays, and possibly $\gamma$-rays as on the Sun: a
striking example of a fast multi-wavelength phenomenon that
requires really simultaneous observations in different bands.

A flare is generally characterized by an unpredictable flux
increase of the order of up to 100 times the quiescent flux in a
typical time scale of the order of 10-100 s. Then the flux
decreases to the pre-flare level on time scales 10-100 times
longer (see Rodon\'o 1990). The possibility of observing
$\gamma$-ray emission from stellar flares has been predicted,
among others, by Becker \& Morrison (1974). Multi-wavelength
studies (Rodon\'o \& Cutispoto 1988; Rodon\'o et at. 1989) have
shown that on the occasion of intense flares, with enhanced
microwave, UV and optical flux increase, the IR flux decreases by
a few percent. Such ''negative'' flares are predicted by a
non-thermal model based on the inverse Compton process (Gurzadian
1977) and a thermal model based on the increased opacity of the
negative H ion (Grinin 1976).

However, owing to the paucity of the observed events, no
conclusion about their nature can be drawn. What is impressive is
that the missing energy in the K-band alone can account for the
energy flux increase at all other wavelengths. Relatively small
''negative'' flares or flare ''dips'' are sometimes observed close
to the occurrence of some flares (Rodon\'o et al. 1979, Cristaldi
et al. 1980).

\subsubsection{Serendipity and Targeted Observations}

In addition to primary and secondary targets a number of
by-products are expected from REM observations. These are
serendipitous discoveries of variable objects falling in the
transients field and precise determinations of NIR colors and
continuum shape of any object falling in the same fields.

The processing and storage of these serendipitous data will be
done off-line and based on the storage media transported to the
REM European Headquarters. There instrumental magnitudes will be
computed and converted into standard colors. There as well the
correlation between images (and spectra) collected at different
epochs will allow one to disentangle variable phenomena and, if
the case analyze their characteristics. All the data will be made
public in a short time in a data base accessible via the internet.

A limited amount of time will also be available for individual
proposals. The handling of this time will be taken care of by the
REM science team which will receive, process, accept or refuse the
proposal. The REM team will then define the procedure and
scheduling of such individual observations and provide the
applicant with the data.

\section{The REM telescope}

The REM telescope is a Ritchey-Chretien system with a 60 cm f/2.2
primary and a overall f/8 focal ratio mounted in an alt-azimuth
mount in order to provide stable Nasmyth focal stations, suitable
for fast motions. REM has two Nasmyth focal stations although at
first light one will remain idle. At the first focal station a
dichroic, working at 45 degrees in the f/8 convergent beam, will
split the beam to feed the two first light instruments of the REM
telescope: the REM-IR camera and the ROSS Spectrograph. The
mirrors are made by Zeiss and they are coated with protected
silver to  maximize reflection efficiency in such a larege (0.45 -
2.3 $\mu$) wavelength range.

The dichroic, a crucial element, has been designed in house with
the aim to achieve the maximum possible transmission at IR
wavelengths (0.95-2.3 $\mu$m) and the maximum possible reflection
at Optical wavelength (0.45-0.95 $\mu$m). Unavoidable losses will
occur in the transition region, i.e the region in which the
efficiency of both transmission and reflection drops. Care has
been then taken to define where the cut should be positioned
according to the scientific requirements and according to the
response at shorter and longer wavelengths of the visible and the
IR detector. The solution adopted provides a cut at 0.95 $\mu$m
and is made of a multi-layer coating of MgF$_2$ and ZnSe on a
substrate of IR-SiO$_2$. The transmission curve of the dichroic is
given in figure \ref{fig9}.

The reflectivity  reported in figure \ref{fig9} is the average
between that computed at 42$^o$ e 48$^o$. The transmission of the
device is strictly complementary to the reflectivity since the
materials used have negligible absorption. The average
reflectivity between 0.45 and 0.90 $\mu$m is 0.965 with a minimum
of 0.880. The average transmission between 1 and 2.3 $\mu$m is
0.972 with a minimum of 0.91. There is obviously a polarizing
effect (the curves are the mean of the S and P components) that
will have to be calibrated once the device is mounted at the
telescope.

\begin{figure}
   \resizebox{\hsize}{!}
   {\includegraphics{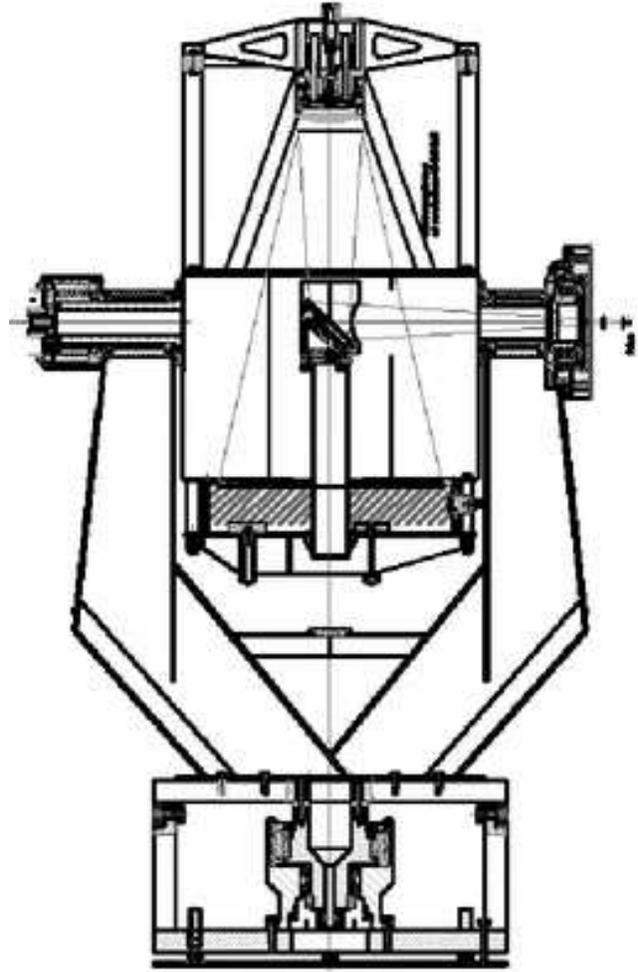}}
  \caption{Schematic overview of the telescope mechanics. Courtesy of
  Teleskoptechnik Halfmann.}
  \label{fig5}
\end{figure}

The REM telescope is currently being manufactured by Halfmann
Teleskoptechnik GmbH (HT) in Neus\"a$\beta$ (Augsburg, BRD). HT
has expertise in the field of professional astronomical
instrumentation. Engineers from HT  actively took part in the
design phase of the telescope suggesting the solution best
suitable to our requirements. A schematic layout of the telescope
mechanics is reported in figure~\ref{fig5}

\subsection{Telescope Control and Operation}

The robotic character of the REM telescope and the minimization of
the need for human intervention is mainly obtained via a robust
operation scheme and a highly modular and reliable operation
software. The main aim that drives the project is to deliver a
system able to take decisions in a few seconds without any human
intervention. However, an adequate interface to allow a remote
human control of operations in case of necessity is developed. In
principle, our system can be subdivided from a logical point of
view in several subsystems that manage the various duties of the
experiment: the REM Observing Software (REMOS), the REM Telescope
Control System (REMTCS), the REM Camera Control Software (REMCCS),
the REM Dome Control Software (REMDCS), the REM Environmental
Control Software (ECS) and the ROSS spectrograph control software.
All these subsystems can work on different or common CPUs,
depending on the specific needs. A certain degree of redundancy is
foreseen.

The most general working scenario starts with the reception by an
efficient socket connection of an alert message from the GCN
(http://gcn.gsfc.nasa.gov/gcn/). The message announces the
detection of a GRB. Now, the REMOS checks if the possible target
satisfies some minimum visibility constraints (hour angle,
altitude, Sun and Moon distance, etc.). If these conditions are
verified, information from REMECS is retrieved and if observations
are possible (humidity, wind, sun altitude, etc., within safety
and operational limits) REMOS checks if the system is already busy
with observations of a different target. If this is the case, and
the source is another GRB, a decisional algorithm is applied.
Depending on the specific case (Infrared Transient, IT, already
discovered, $\gamma$-ray fluence, favorable sky position, time
interval from the GRB, etc. ) eventually a decision is taken. In
the positive case (new GRB to be observed) or in case the
telescope was observing some lower priority target , REMOS sends a
message via socket connection the REMTCS stopping operations and
moving the telescope to the new target. As soon as the REMTCS
communicates that the telescope is on target the REMOS sends
messages to the REMCCS and ROSS to be-gin observations with some
predefined templates. Immediately after the first frames are
obtained, they are analyzed by the REM Reduction Software (REMRS)
that can, if necessary, send new coordinates to the REMOS, i.e. to
better center the target, or modify the observing templates, i.e.
in case of peculiarly bright or faint IT.

Of course the REMOS also periodically samples the environmental
conditions in order to drive the opening or closing of the dome by
the REMDCS and stops operation in case some safety alarm is
triggered (un-expected intrusion in dome, etc.). Apart from the
reaction to a GRB alert, it is possible by a web based interface
to provide the telescope coordinate list and observing parameters
for any target to be observed for the additional science programs.
These operations follow basically the same outline, even if any
new low priority target is observed only after the completion of
all operations on the previous one on the priority list. It is
also possible to prepare observations of a target on a specific
schedule, i.e. for simultaneous observations, or under specific
observing conditions, i.e. zenith distance.

\section{The REM-IR Camera}

\begin{figure}
   \resizebox{\hsize}{!}
   {\includegraphics{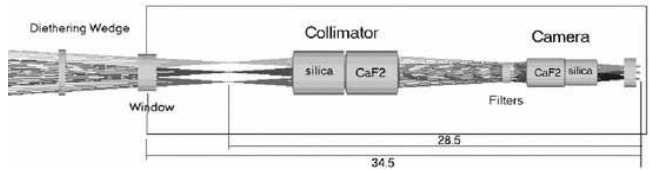}}
  \caption{REM-IR camera design. The camera is a focal reducer (F/8-
F/5.3) with separated doublets for collimator and camera. A cold
stop is inserted for background reduction where the pupil is
reformed: the filters are as well inserted in the collimated beam.
In front of the cryostat window we find the rotating plate used
for dithering}
  \label{fig6}
\end{figure}

\begin{figure}
   \resizebox{\hsize}{!}
   {\includegraphics{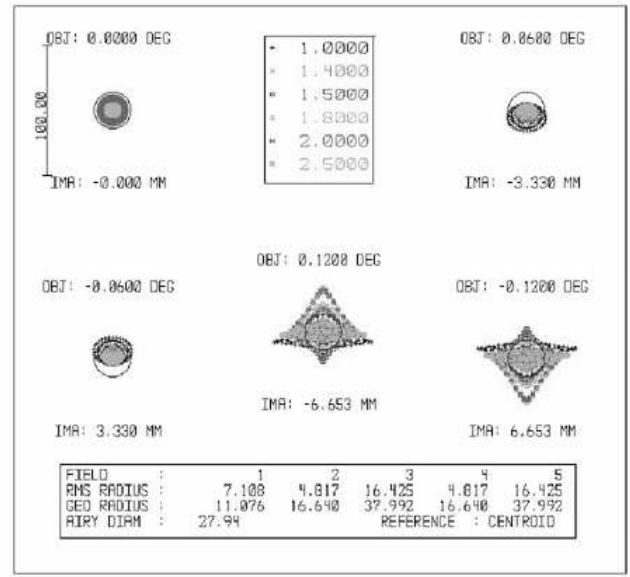}}
  \caption{Polychromatic (1.0 - 2.5 $\mu$m) spot diagrams for REM-IR
Camera. The bold circle represent the diffraction limit which is
reached almost at any field.}
  \label{fig7}
\end{figure}

The REM camera follows a focal reducer design in order to reform a
white pupil in a cold environment for Lyot-stop positioning (see
figure~\ref{fig6}). A filter wheel with 10 positions is located at
the reformed pupil allowing one to insert filters and grisms for
slit-less spectroscopy or polarimeters in a parallel beam. The
camera changes the focal ratio from f/8 to f/5.3 providing a
plate-scale of 64.4 as/mm that allows one to position a 9.9 x 9.9
am$^2$ FOV on a 512x512 (18 $\mu$m pitch) HgCdTe chip in
production at Rockwell. Both collimator and camera are made of a
Silica-CaF2 doublet (the latter reverse-mounted). The total
thickness of the optical design is 300 mm The image quality is
optimal as can be seen in the spot diagrams reported in
figure~\ref{fig7}

\begin{figure}
   \resizebox{\hsize}{!}
   {\includegraphics{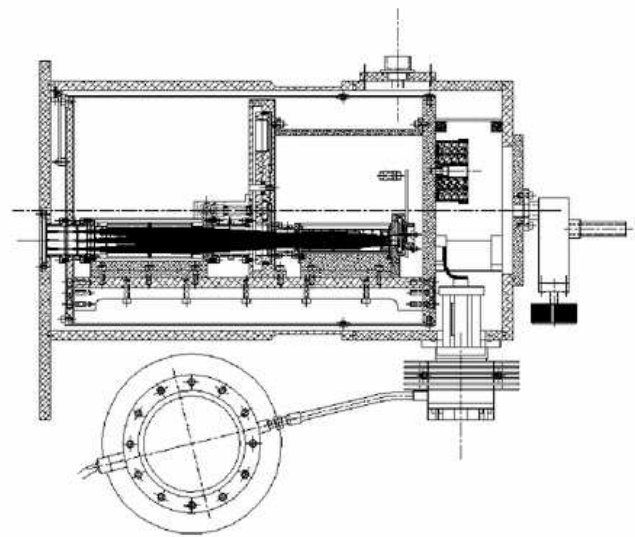}}
  \caption{Mechanical layout of the REM-IR Camera. The optics are
mounted in slightly decentered vacuum dewar. The detector, the
cold stop and the inner lenses are cooled via a closed circuit
Stirling Cryo-pump. Drawing courtesy of IRLabs, Arizona }
  \label{fig8}
\end{figure}

The whole camera train is mounted in a dewar and operated in a
cool environment. The chip working temperature is 77 K and will be
guaranteed at the detector location and at the cold stop position.
The optical train is kept at a temperature of about 100-120 K in
order to save cooling power. The cryogenics are supported by a
Stirling-Cycle cryo-pump requiring limited maintenance and no need
for dewar refilling. A sketch of the mechanical layout of REM-IR
camera is shown in figure~\ref{fig8}.

The REM-IR camera needs dithering of the images for noise
reduction and sky subtraction. However such a dithering can not be
provided moving the telescope since the ROSS spectrograph needs a
stable image for longer integrations. The diethering will then be
provided by a plate with plane parallel surfaces inserted with a
given tilt (18$^o$ ) in front of the cryostat window, i.e. after
the dichroic. The effect of the plate is to translate the image of
20 pixels from the nominal position (no plate inserted) along a
given direction. By rotating the plate in the beam such a
direction and hence the displacement of the image can be changed
at will providing the dithering. A rotary stage changes the
orientation of the plate, the angle is read by an encoder and
transmitted to the Reduction software making the image
reconstruction possible.

\begin{figure}
   \resizebox{\hsize}{!}
   {\includegraphics{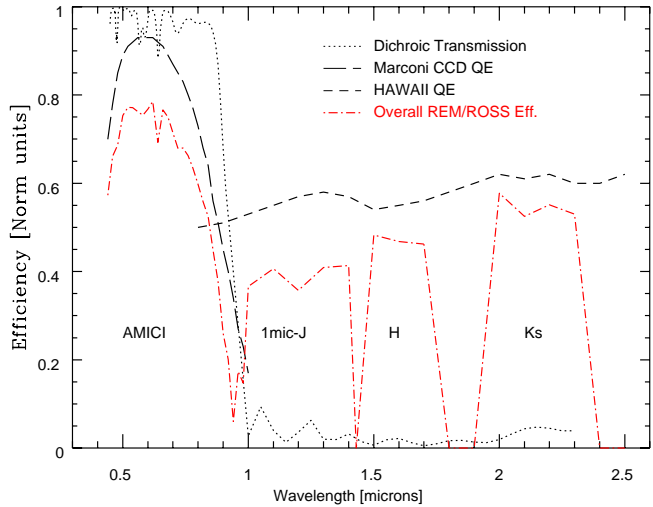}}
  \caption{Overall Efficiency of the REM IR camera and ROSS. We show
in the plot the transmission of the dichroic, the QE of the
Marconi CCD and of the HAWAII Chip and an estimate of the
''all-included'' efficiency of the instruments.}
  \label{fig9}
\end{figure}

By taking into account glass transmission and reflection (with
suitable AR coating) the overall transmission of the camera
(telescope, array and filters not included) is estimated to be
about 85\%. An additional 2\% loss should be expected from the
Telescope (silver coated) and the minimum expected transmission of
the dichroic (see above) is 91\% . HAWAII chips have peak
efficiency around 64\% while from 0.9 to 2.5 $\mu$m it is never
lower than 56\%. Therefore we should expect an overall efficiency
of about 59\% (43\% minimum; filters excluded). The overall
efficiency of the REM-IR camera is reported in figure~\ref{fig9}

The regime in which REM-IR is operated is mainly sky limited and
RON and DARK have little effect on the performance. We computed
the limiting magnitudes using the usual formula for S/N
calculations and NTT-SOFI J,H and Ks filters as a template for REM
filters still under procurement. We assume the following average
transmissions: 75\%, 80\% and 85\% for J,H,K' respectively. No
predictions have been made for the Z' filter since its
transmission will be modulated  a posteriori on the cut-on of the
HgCdTe chip and the dichroic. The limiting magnitudes computed
using these figures are reported in table 1.

\begin{table}
\begin{center}
\caption{Limiting magnitudes as a function of the S/N (10 or 5)
and of the passband, for the different integration times foreseen
in the REM target operation.} \vspace{0.5cm}
\begin{tabular}{|l|c|c|c|c|c|c|}
\hline \hline T int. & J & J & H & H & K & K\\
      &10$\sigma$&  5$\sigma$&   10$\sigma$&  5$\sigma$&   10$\sigma$&  5$\sigma$ \\
\hline 5 sec &   15.5  &  16.3  &  14.3 &   15.1 & 13.1 & 13.9
\\
30 sec & 16.4  &  17.2  &  15.2  &  16.0  &  14.1  & 14.8
\\
600 sec &17.3  &  18.0 &   16.3  &  17.1  &  15.4  & 16.1
\\ \hline \hline
\end{tabular}
\end{center}
\end{table}

The REM selected FOV has been selected to match the typical
error-boxes of $\gamma$-ray alert systems. The need of absolute
photometric measurements (magnitudes and IR color indices) makes
the presence in the FOV of a number of comparison non-variable
stars necessary. Simulation using the USNO-A1.0 Catalogue of
objects from digitized plates have been performed. Using the
server at the ESO/ST-ECF Archive an automated procedure retrieves
a star chart showing the magnitudes and V-R colors of the stars.
Targets are of random fields or for a given coordinates. Random
fields are used to build a luminosity function in the FOV and
preliminary results for the search of R=15 stars ranges from null
detection to several dozen of objects. Part of the idle time of
the REM telescope (i.e. on clear nights when the satellite is
pointing to a forbidden, below horizon area) will be partly
devoted to post-calibration of observed fields.

\subsection{The Automatic Quick-Analysis Software}

The Automatic Quick-Analysis software (AQuA) has been developed
and entirely dedicated to the REM data using criteria of high
speed, system stability and reliable results in a fully automated
way. It will run on a high performance computer with double
processor completely dedicated to data handling in order to find
transient coordinates and colors. Time sensitive data are quickly
distributed via internet to the recipients entitled to receive
them (including ToO procedure at larger telescopes) while the bulk
of data (2.5 Gbytes of data per night expected) is recorded in a
portable storage system. In this scheme the only human
intervention needed during normal operation is to change such a
removable science team approximately once every 10 days.

Given the intrinsic rapidity of the phenomena rapid photometry is
as essential as the rapid mount drive. Based on a flux model which
evolves as $f\sim t^{-\delta}$ with $0.8<\delta<2.0$ we reckon
that a measurement of each filter Z', J, H and K every 5 seconds
in the first observations after targeting is the minimum
acceptable frequency. Slower data flow will be allowed after a 1.5
minutes and a measurement every 30 seconds will be delivered. One
hour after the trigger signal the frequency will be definitely
slowed down to one measure every 10 minutes. We judge that,
focussing on targets with a typical known visual (V) magnitude
between 9 and 15, the system response is within scientific
specifications if no abrupt changes in the spectral energy
distribution show up.

The system is normally working on secondary science activity and
waiting for a trigger from the camera acquiring system. When the
trigger arrives the system is ready to receive FITS files from the
camera directly on a shared disk. It has no command duty with the
exception of a line open with the camera control to refine
exposure parameters and provide refined coordinates. When the
telescope starts to acquire the target the data analysis system
will be alerted and it will wait for the first set of images.  As
the first set of five dithered images will be available on the
disk they will be preprocessed in order to obtain one scientific
image corrected for bad pixels, flat and sky subtracted. As soon
as the final image is ready it will be analyzed by a source
detection alghoritm and a list of targets will be extracted.
Source positions will be compared with catalogues to both perform
astrometry and look for possible candidates. In the following step
photometry will be performed on the field and calibrated with
2MASS sources (if present in the field) or using instrumental
characteristic (e.g. exposure time vs limiting magnitude). AQuA
has been prepared and optimized in order to perform all these
operations in few seconds, about the same time needed for the
acquiring system to collect another set of images. When a second
scientific image becomes available a second source list will be
extracted and compared with the previous one looking for variable
sources. Furthermore the two images will be subtracted and
filtered and the resulting image will be inspected with the
detection algorithm and results compared with the previous
results. If any source is present in the field the Automatic
Quik-Analysis Software will provide the Observation Software with
a warning and a higher exposure time. On the contrary if a
transient source will be found the AQuA will provide the OSW with
the coordinates.  Coordinates will be then delivered in an
automatic way via e-mail to already existent Alert Networks (e.g.
GCN) and/or to a dedicated mailing list (e.g. REM Alert Mail).As
soon as positions from the SWIFT optical monitor will be delivered
AQuA will cross check them with REM positions. If an IR transient
is found but no optical transient is present in the field, AQuA
will activate a ToO to the VLT providing it with transient
coordinates. Once the IR source has been found the OSW will change
the filter and AQuA will perform the analysis on the other images
collecting magnitudes in different bands and performing color
measures.

Consequently when operated in the Primary science mode, i.e.
following GRB triggers, REM will produce 3 classes of output on
different time-scales: the coordinate of the IR transient will be
available in a few seconds; magnitudes and colors will be
available in a few minutes (possibly giving a rough estimate of
the distance of the burst via the photometric red-shift
technique); light and color curves will be accurately computed
off-line by an extended version of AQuA. Coordinates and colors,
expecially in the case of a high-redshift burst, are probably the
most valuable science output REM can provide.

\section{The ROSS Spectrograph}

With an orthogonal development relative to the REM-IR camera
optical axis REM Nasmyth A will accommodate the slitless
spectrograph ROSS. The spectrograph consists of a fore optics
which images a pupil at the location of the dispersing element and
re-maps the focal plane onto the detector unit. The selected
detector head is a commercial Apogee AP47 camera hosting a Marconi
47-10 1K x 1K 13 $\mu$m pitch CCD. The plate scale of the REM
telescope (43 arcsec/mm) matches properly with the specifications
and allows one to cover a 9.54 X 9.54 am$^2$ with a scale of 0.56
as/px. The fore-optics have been then designed at magnification 1.

\begin{figure}
   \resizebox{\hsize}{!}
   {\includegraphics{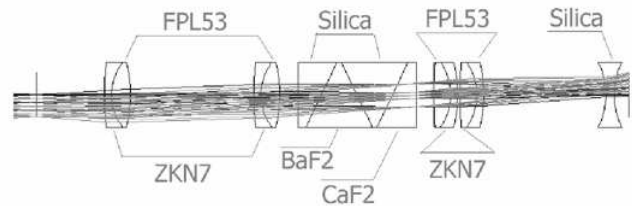}}
  \caption{ROSS Optical Layout. ROSS has a focal reducer scheme to
  reform a pupil to accommodate the AMICI prism. The natural scale of the
telescope is however well suited for this instrument: ROSS works
at magnification 1 }
  \label{fig10}
\end{figure}

The optical layout of the spectrograph is reported in
figure~\ref{fig10}. The collimator is made of a pair of identical
ZKN7- FPL53 separated doublets while the camera is made of two
identical FPL53-ZKN7 separated doublets: the pairs differs one
from the other. A SILICA window lens with identical curvature on
both sides (easier mounting) closes the Apogee detector head.

\begin{figure}
   \resizebox{\hsize}{!}
   {\includegraphics{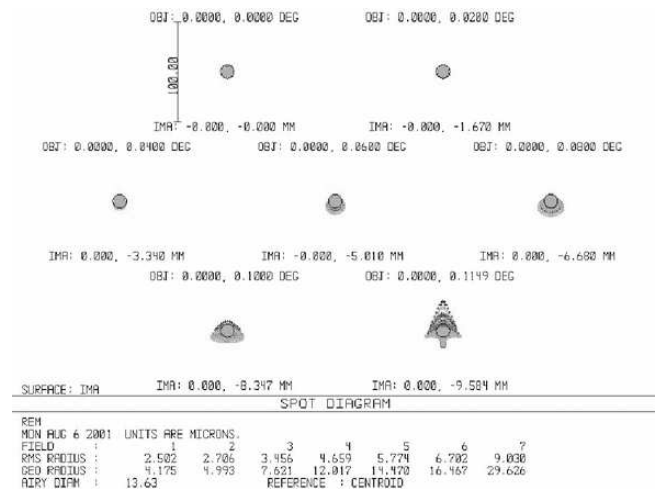}}
  \caption{ROSS Spot Diagram (with no AMICI inserted in the beam).
The bold circle represents the diffraction limit which is reached
for almost any field.}
  \label{fig11}
\end{figure}

The dispersion is obtained by insertion at the pupil location of
an Amici Prism 66 mm long. The prism is made of Silica, BAF2 and
CAF2 and it spreads the 0.45-0.95 $\mu$m wavelength range on 60
pixels, allowing the recording of 30 2-pixels bins along the
range. The optical quality of ROSS is good as can be seen in the
spot diagram reported in figure~\ref{fig11}

By taking into account glass transmission and reflection (with
suitable AR coating) the overall transmission of the camera
(telescope, dichroic, CCD and AMICI prism not included) is
estimated to be about 85\%. An additional 2\% loss should be
expected from the telescope (silver coated), the minimum expected
reflection of the dichroic (see above) is 88\% and the AMICI is
exected to uniformly lose about 2\% at any wavelength. With the
above numbers and the MARCONI CCD QE curve we computed the overall
efficiency of ROSS; the curve is reported in figure \ref{fig9}.

\begin{figure}
   \resizebox{\hsize}{!}
   {\includegraphics{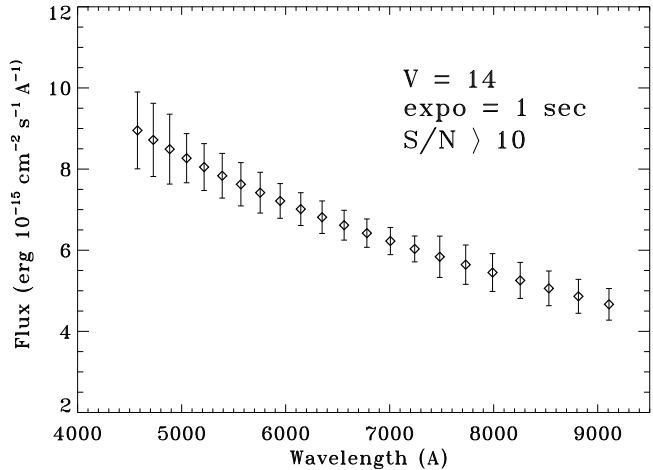}}
  \caption{Simulated spectrum of a magnitude V=14 GRB optical
flash displaying a spectral slope $F_\lambda \propto \lambda
^{-1.25}$. The exposure time is 1 second. }
  \label{fig12}
\end{figure}

In order to evaluate the expected performance of the ROSS
spectrograph a simulated spectrum of a V=14 GRB optical flash with
a spectral slope $F_\lambda \propto \lambda ^{-1.25}$ was
computed. The spectrum for a 1 sec exposure is shown in
figure~\ref{fig12} where we can see that a $S/N>10$ is reached at
any wavelength.

\begin{figure}
   \resizebox{\hsize}{!}
   {\includegraphics{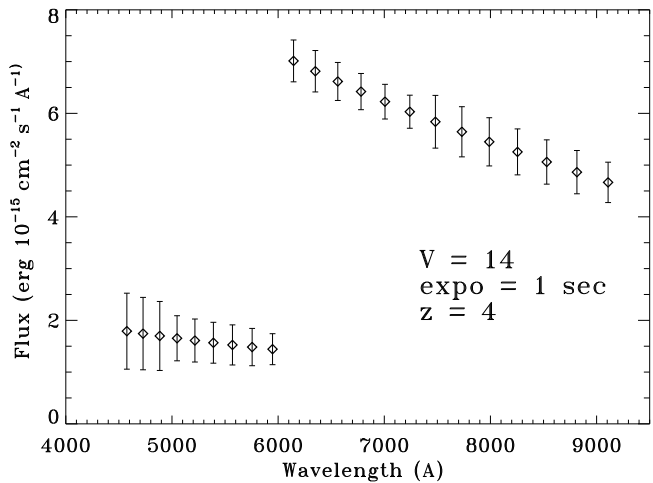}}
  \caption{Simulated spectrum of a magnitude V=14 GRB optical flash
displaying a spectral slope $F_\lambda \propto \lambda ^{-1.25}$
and a $Lyman-\alpha$ dropout which falls in the optical range due
to the high redshift ($z$) of the source. The exposure time is 1
second and the signal to noise ratio is about 4 below the break
and greater than 10 above it.}
  \label{fig13}
\end{figure}

By inserting a Ly-$\alpha$ drop due to high redshift of the source
we obtain the spectrum shown in figure~\ref{fig13} where the S/N
below the break is of about 4.

\section{Site Location and Arrangements}

A telescope like REM, specially because of the IR camera, has to
be located in a dry and high altitude site with good weather
conditions and a high percentage of observing nights. A viable and
reliable connection to a large facility accepting Target of
Opportunity alerts is in addition an important issue for the
success of the project. For this reason we explored possible
agreements with exisiting Observatories with a natural preference
for the European Southern Observatory site at la Silla and Cerro
Paranal, Chile.

\begin{figure}
   \resizebox{\hsize}{!}
   {\includegraphics{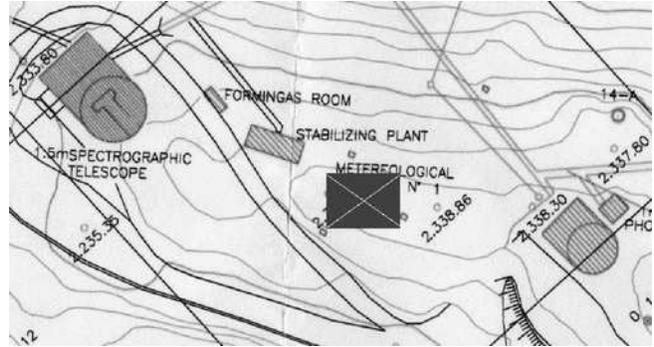}}
  \caption{The site where the REM observatory will be built, in between
ESO 1.52 and ESO 1.0 telescopes at la Silla Observatory, Chile.}
  \label{fig14}
\end{figure}

An agreement to install REM at la Silla Observatory has been
achieved in the framework of the FROST (Fast Robotic Observatory
System for Transient), formed by REM and the French Project
TAROT-S (see Boer et al, these proceedings). REM will be installed
at la Silla Observatory at UTM (Zone 19) E 331,235 N 6,762,735
elev. 2,338.80 mt (see the exact location in the map in
figure~\ref{fig14}

The observatory building is currently under construction. The
Telescope will see the first light in October 2002.

\end{document}